\documentclass[noshowpacs,twocolumn,aps,pre]{revtex4}
\usepackage[dvips]{graphicx}
\usepackage{amssymb}
\usepackage{amsmath}
\usepackage{latexsym}
\usepackage{epsfig}
\usepackage{bm}
\usepackage{times,psfrag,subfigure}        
\begin{document}

\title{Anisotropic hysteresis on ratcheted superhydrophobic surfaces}
\author{H. Kusumaatmaja${}^{a,b}$} 
\author{J. M. Yeomans${}^{a}$}
\affiliation{${}^{a}$The Rudolf Peierls Centre for Theoretical Physics, Oxford University, 1 Keble Road, Oxford OX1 3NP, U.K.}
\affiliation{${}^{b}$Max Planck Institute of Colloids and Interfaces, Science Park Golm, 14424 Potsdam, Germany}
\date{\today}

%%%%%%%%%%%%%%%%%%%%%%%%%%%%%%%%%%%%%%%%%%%%%%%%%%%%%%%%%%%%%%%%%%%%%%%%%%%%%

\begin{abstract}
We consider the equilibrium behaviour and dynamics of liquid drops on a superhydrophobic  
surface patterned with sawtooth ridges or posts. Due to the anisotropic geometry of the  
surface patterning, the  contact line can preferentially depin from one side of the  
ratchets, leading to a novel, partially suspended, superhydrophobic state. In both this  
configuration, and the collapsed state, the drops show strong directional 
contact angle hysteresis as they are pushed across the surface. The easy direction is,  
however, different for the two states. This observation allows us to interpret recent  
experiments describing the motion of water drops on butterfly wings.
\end{abstract}
%\pacs{68.08.Bc, 47.61.Jd, 47.61.-k}
\maketitle

%%%%%%%%%%%%%%%%%%%%%%%%%%%%%%%%%%%%%%%%%%%%%%%%%%%%%%%%%%%%%%%%%%%%%%%%%%%%%

Superhydrophobic surfaces  \cite{Quere} show extreme water repellency, with  
effective contact angles that can be close to $180^\circ$. Superhydrophobicity results  
when the natural hydrophobicity of a substrate is amplified by roughness. Typically a  
drop on a superhydrophobic surface can either lie in a Cassie-Baxter state  
\cite{Cassie}, suspended on top of the roughness with air-pockets beneath, or in the  
Wenzel state \cite{Wenzel}, where the liquid penetrates the spaces between the surface  
corrugations. The drop contact angle is enhanced in both the Wenzel and the  
Cassie-Baxter states but the dynamical behaviour of the two configurations is very  
different; a liquid drop in the suspended state is highly mobile, while that in the  
collapsed state is strongly pinned.

Superhydrophobic surfaces have evolved naturally in many contexts. Plants, such as the  
lotus, nastertium and Lady's Mantle have superhydrophobic leaves to aid the run-off of  
rain \cite{Neinhuis}. The plasteron used by aquatic insects to breathe underwater is air  
trapped beneath a Cassie-Baxter interface, and hairs on the legs of water insects render  
them superhydrophobic and lead to enhanced buoyancy \cite{Bush}. In a recent paper Zheng  
{\it et. al.} \cite{Zheng} reported that butterfly wings are hydrophobic and that they  
are patterned by an arrangement of anisotropic scales of typical size 100$\mu m$. These  
authors found that liquid drops placed on the wings roll easily away from the insect's  
body, but are pinned strongly against movement in the opposite direction, towards the  
body. 

\begin{figure} 
\begin{center}
\includegraphics[scale=0.5,angle=0]{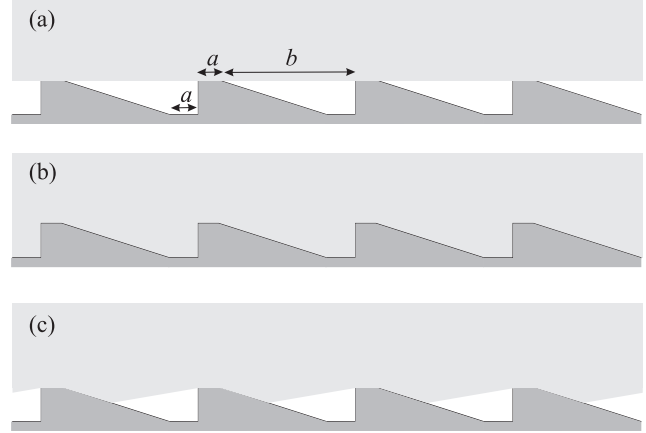}
\caption{Schematic diagram of a drop on a hydrophobic surface patterned with ratchets.  
(a) Suspended, (b) collapsed and (c) partially suspended state.} \label{schematic}
\end{center}
\end{figure}

Motivated by this work we investigate the behaviour of drops on a hydrophobic surface  
patterned with ratchets, posts which have sides of different slopes, see  
Fig.~\ref{schematic}. We identify a novel, partially suspended, superhydrophobic state  
and describe its regions of stability. We then simulate the hydrodynamics of drops  
pushed across the ratchets. We find that the partially suspended and collapsed  
configurations both have a highly anisotropic response to a driving force, effectively  
functioning as `fluidic diodes'. Surprisingly, the easy direction is opposite for the  
two configurations, primarily due to the behaviour of the receding contact line. The new  
results enable us to explain the motion of raindrops on butterfly wings.

Consider a large liquid drop lying on a regular array of hydrophobic ridges, one side of  
which makes an  angle $\alpha$ with the horizontal, and the other side of which is  
vertical, as shown in Fig.~\ref{schematic}.   Figs.~\ref{schematic}(a) and (b) depict  
the Cassie-Baxter, or suspended, and Wenzel, or collapsed, configurations respectively.  
On a ratcheted surface, however, a third configuration is possible. We shall term this,  
shown in Fig.~\ref{schematic}(c), the partially suspended state. 

The partially suspended state is stable because, when a drop is placed gently 
onto a ratcheted surface, the Gibb's criterion \cite{Gibbs} 
states that the interface will remain pinned at the ridge corners if the angle between 
the interface and the sides of the ratchets is less than the intrinsic equilibrium contact 
angle of the flat surface $\theta_e$  \cite{Kusumaatmaja,Tuteja}. For strongly  
superhydrophobic surfaces and large values of $\alpha$ this criterion is satisfied and  
the drop remains in the Cassie-Baxter state. As the wettability of the surface is  
increased, or $\alpha$ is decreased, depinning will occur. This happens first on the  
corner abutting the ridge side with the smallest gradient. For a drop, large compared to  
the size of the ridges, but small enough that gravity can be neglected, the condition  
for depinning is (for the two-dimensional, ridged geometry)
\begin{equation}
\theta_e < \pi-\alpha \, . \label{tr1}
\end{equation}
The transition is reversible: the contact line moves slowly down the sloping side of the  
ratchets as $\theta_e$ is decreased from the threshold value, and the interface will  
move back to the corner if the contact angle is increased above the value given in  
Eq.~(\ref{tr1}).

As the equilibrium contact angle or $\alpha$ is decreased further the drop depins from  
the second corner and moves down the vertical side of the ridges thus collapsing to the  
Wenzel state. For a large drop this happens when
\begin{equation}
\theta_e = \frac{3\pi}{4}-\frac{\alpha}{2} \, . \label{tr2}
\end{equation}
Once depinning has occurred the interface moves immediately to the base of the grooves.  
This transition is irreversible, because the liquid--gas and gas--solid interfaces are  
replaced by a single liquid--solid interface, and there is an energy barrier to  
re-forming the gas layer.

\begin{figure} 
\begin{center}
\includegraphics[scale=0.81,angle=0]{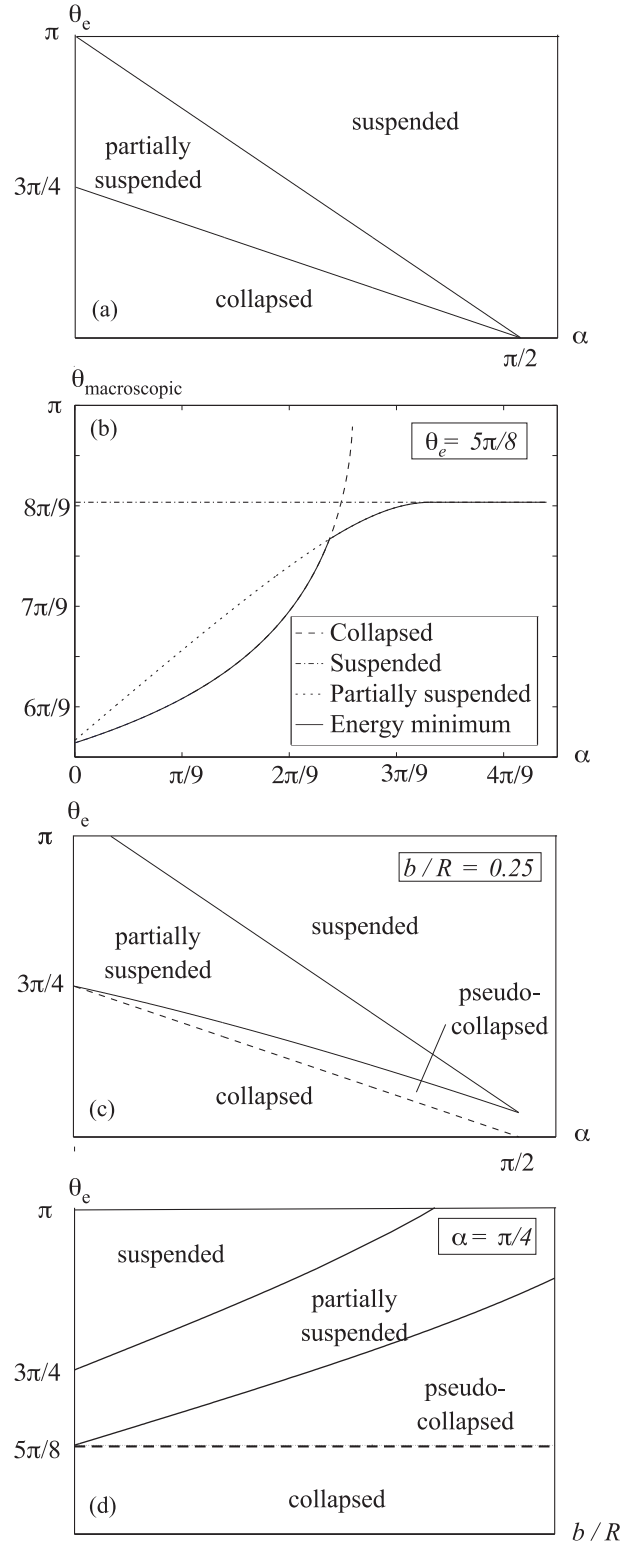}
\caption{(a) Regions of stability of the suspended, partially suspended and collapsed  
states of an infinite drop placed gently on a two-dimensional ratcheted surface. (b) Variation of the macroscopic contact angle with the ratchet angle $\alpha$ for  
$\theta_e=5\pi/8$ and $a/b=0.1$. (c) as (a), but for a finite drop, $b/R=0.25$, (d)  
Regions of stability of the different states, for varying contact angle and drop size,  
for $\alpha=\pi/4$.}  
\label{phdiagram}
\end{center}
\end{figure}

The regions of stability of the different states, for an infinite drop initially placed  
on top of the ratchets, are summarised in Fig.~\ref{phdiagram}(a). For $a=0$, the figure  
also represents the phase diagram, ie the global minimum of the free energy.

Cassie-Baxter and Wenzel used simple thermodynamic arguments to derive the effective  
macroscopic (observable) contact angle for drops on superhydrophobic surfaces. Their  
formulae are applicable for drops covering a large number of ridges, which are not  
pinned with respect to motion across the surface. For the geometry we consider,  
arguments of this sort give a macroscopic contact angle
\begin{equation}
(a+b) \cos(\theta_{C-B}) = a \cos(\theta_e) - b, 
\end{equation}
\begin{equation}
(a+b) \cos(\theta_{PS}) = a \cos(\theta_e) + b \cos(\theta_e + \alpha), 
\end{equation}
\begin{equation}
(a+b) \cos(\theta_{W}) = \left(2a + \frac{(b-a)(1+\sin{\alpha})}{\cos{\alpha}}\right)  
\cos(\theta_e)
\end{equation}
for the Cassie-Baxter, partially suspended, and Wenzel state respectively.  
Fig.~\ref{phdiagram}(b) compares the macroscopic angles of the three states as a  
function of $\alpha$ for $\theta = 5\pi/8$ and $a/b = 0.1$. The state with the lowest  
contact angle corresponds to the minimum energy configuration \cite{Okumura,Patankar}.

We now discuss finite drops and the way in which changes in drop radius of curvature, due to, say,  
evaporation or application of external pressure, can lead to transitions. For finite drop radius the curvature of the drop  
makes it easier for it to depin from the corners of the ratchets and the  
transition from the suspended to partially suspended state occurs at
\begin{equation}
Rc_1 = \frac{b}{2\sin(\theta_e+\alpha-\pi)}. 
\end{equation}
Indeed for sufficiently small $\alpha$,  $\theta_e$ and $R$ the suspended state is never  
stable, and the drop falls immediately into the partially suspended state. As the drop  
size is deceased further depinning from the second corner of the ratchet occurs at
\begin{equation}
Rc_2 = \frac{b\tan(\alpha)\sin(\pi/4-\alpha/2)}{\sin(\theta_e+\alpha/2-3\pi/4)}.
\end{equation}
These boundaries are shown in Fig. \ref{phdiagram}(c) as a function of $\theta_e$ and  
$\alpha$ for $R/b=0.25$ and in Fig. \ref{phdiagram}(d) as a function of $\theta_e$ and  
$b/R$ for $\alpha=\pi/4$.

For an infinite drop the depinned interface moves immediately to the bottom of the  
groove. However this is no longer the case for a finite drop. Because of the ``V'' shape  
geometry, the pressure set up by the sides of the grooves (because the local contact  
angle is equal to the Young's angle in equilibrium) increases as the drop penetrates  
them. This means that the interface moves slowly down the ridges as the drop volume  
decreases: there is  a stable interface profile for a given drop volume, or  
equivalently, for a given external pressure \cite{Brinkmann}. We shall term this the `pseudo-collapsed' state as its  
dynamical properties are similar to those of a Wenzel configuration. The transition  
from partially suspended to pseudo-collapsed is reversible, the contact line moves  
slowly up and down the sides of the ratchets as a function of the Laplace pressure. The  
drop remains in the pseudo-collapsed configuration until the condition given in  
Eq.~(\ref{tr2}) is satisfied, at which point the interface slides immediately to the  
bottom of the groove. This transition is irreversible. Fig.~\ref{phdiagram}(c) shows the  
configurations which are global minima of the free energy for $a=0$ and $b/R=0.25$. For  
$a \ne 0$ the sequence of transitions is truncated by a discontinuous change to the  
Wenzel configuration immediately the interface touches the base of the grooves.

For a drop on posts, rather than ridges, it is not possible to describe the transitions  
analytically, because of the two-dimensional curvature of the interfaces between the  
posts. However the same quantitative behaviour is expected. Indeed, a sequence of  
depinning transitions could occur on posts with facets of different slopes.
\begin{figure} 
\begin{center}
\includegraphics[scale=0.75,angle=0]{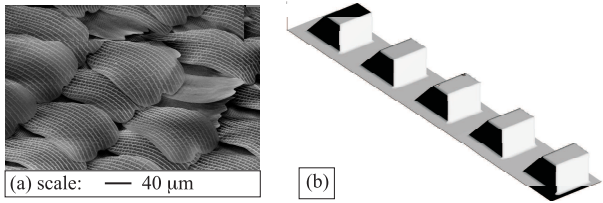}
\caption{(a) The wings of Papilio palinurus. The figure is  
kindly provided by Prof. S. Berthier. (b) The surface patterning used to obtain Fig.  
\ref{Ratchet}.}
\label{geometry}
\end{center}
\end{figure}

We now compare the dynamics of drops in the suspended, partially suspended and collapsed  
configurations, considering a cylindrical liquid drop confined between surfaces  
patterned with posts. This geometry was chosen because it includes the important physics  
of the three dimensional problem, namely the two dimensional curvature of the interface  
between the posts, while being less demanding of computational resources than modelling  
a full three dimensional drop. (A more restricted set of simulations of drops sitting on  
top of a single ratcheted surface gave qualitatively the same behaviour.) Guided by the  
patterning on the wings of Papilio palinurus, shown in Fig.~\ref{geometry}(a), the posts  
were taken to have three vertical and one sloping side (see Fig.~\ref{geometry}(b)). 
On butterfly wings, the sloping angle $\alpha$ is typically of order $10^\circ$ but here,
to restrict the inter-ridge spacing to a computationally feasible value, we have 
used $\alpha=45^\circ$. 
 
The drop thermodynamics was described by a binary free energy model and the   
hydrodynamics by the corresponding Navier-Stokes equations, solved using a lattice  
Boltzmann algorithm. This approach is proving a useful tool to study drop motion in  
situations where pinning and hysteretic effects are important. Details of the model and  
algorithm can be found in \cite{Pooley,HalimThesis} and the references therein. 

\begin{figure} 
\begin{center}
\includegraphics[scale=0.775,angle=0]{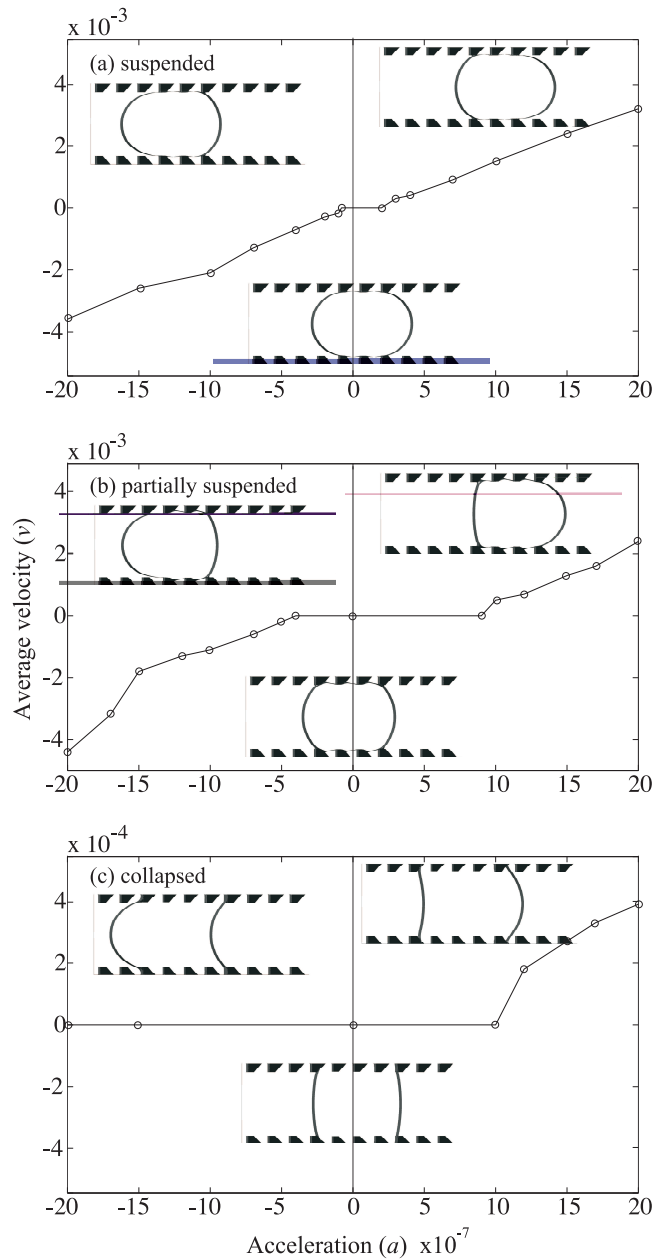}
\caption{Comparison of the mean velocity of (a) suspended ($\theta_e=140^\circ$), (b)  
partially suspended ($\theta_e=110^\circ$) and (c) collapsed  
($\theta_e=100^\circ$) drops as a function of the applied acceleration. Drop shapes in equilibrium  
and when moving in different directions are shown in the insets. The other parameters of  
the simulations are: surface tension $\gamma = 0.013$, liquid drop viscosity  
$\eta_{\mathrm{liquid}}=0.83$, surrounding gas viscosity $\eta_{\mathrm{gas}} = 0.06$.  
All data is in simulation units.}
\label{Ratchet}
\end{center}
\end{figure}

Fig.~\ref{Ratchet} compares the motion of a drop in the suspended, partially suspended  
and collapsed states. The intrisic contact angles for the three states are
$\theta_e = 140^\circ$, $110^\circ$ and $100^\circ$ respectively and the drop covered  
$\sim 4$ posts. The drop was pushed by a variable body force parallel to the surface; we  
choose to refer to the direction from the vertical to the sloping edge of the posts as  
positive (to the right in Fig.~\ref{Ratchet}). 

The average velocity as a function of the applied force is shown in Fig~\ref{Ratchet}.  
There is a striking difference between the three configurations. In the suspended state  
the pinning is weak, and essentially independent of direction. In the collapsed state, however,  
there is strongly anisotropic pinning and the drop does not move at all in the  
negative direction for forces in the range we have considered. This occurs because the  
interface is strongly pinned with respect to movement down the vertical sides of the  
posts \cite{Silberzan}. Once the drop is moving, the velocity for a given force is a factor 
$\sim 10$ lower in the collapsed than in the suspended state.

A pronounced asymmetry in the pinning threshold is also seen for the partially suspended  
state, but note that the negative direction is now the {\em easy} direction.
This surprising behaviour is a result of the shape of the liquid--gas interface  
underneath the drop. From the simulations, we find that the motion of the drop is  
primarily controlled by pinning at the receding contact line. For a partially suspended  
drop pushed in the positive direction, the receding interface lies part-way down the  
posts and the drop dynamics are those of a collapsed state, with strong pinning. 
If the drop is pushed in the negative direction, however, the trailing edge lies on top of the  
posts and the drop behaves like a suspended state, with low contact angle hysteresis
\cite{HalimHysteresis}. 

Butterfly wings form a ratcheted surface, albeit a much more structured one than that we  
consider here. The wings are covered with overlapping scales of length $\sim 100 \mu m$,  
with free ends pointing towards the wing tips (Fig.~\ref{geometry}(a)). Drops of water  
are observed to run away from the butterfly body: Zheng {\it et. al.} \cite{Zheng}  
reported that on the wings of Morpho aega, the pinning threshold is $> 6$ times stronger  
for motion towards the body. 

The easy direction for the butterfly corresponds to the negative direction in our  
simulation geometry, and hence the easy direction for the partially suspended state,  
which should  be stabilised by the anisotropic wing patterning. We do, however, find a  
less strong anisotropy in the simulations, and it will be interesting to investigate  
which model parameters are key in controlling this. We note that a secondary structure  
of ridges of height $\sim 0.5 \mu m$ running along each scale are observed on the butterfly wings.  
This is not needed to contribute to the anisotropic hysteresis, but could be useful in  
preventing lateral run-off or in increasing the effective contact angle of the surface 
\cite{McCarthy}.

Anisotropic surface patterning is not limited to butterfly wings. For 
example, several authors ({\it{e.g.}} \cite{Bush}) have shown that the legs of insects  
such as water striders are covered with hairs that are tilted by $\sim 60^\circ$ from  
the vertical direction. A possible role of the asymmetry is to introduce  
directionality to the motion of the insects on water surfaces.

We have considered a superhydrophobic surface patterned with sawtooth posts and shown  
that a partially suspended configuration can be stable in addition to the two usual,  
suspended and collapsed, phases. It is now feasible to construct ratcheted surfaces on  
micron length scales, so we hope that this letter will motivate experimental work  
identifying the novel configuration on both fabricated and biological surfaces. 

Drops in both the partially suspended and collapsed states have an anisotropic response  
when they are pushed across the surface, but the easy direction is different in the two  
configurations. Furthermore, the transition between the partially suspended and the  
pseudo-collapsed state can be induced reversibly by the changing the drop pressure or  
contact angle (which can be done via electrowetting). This opens up the possibility of  
constructing a `fluidic diode', with external control of the easy direction, for use in  
microfluidic devices.
 
We thank S. Brewer, M. Blow, K. Hermans, R. Lipowsky, and R. Vrancken for useful  
discussions. We gratefully acknowledge Prof. S. Berthier (Institut des Nano-Sciences de Paris)
for providing Fig. \ref{geometry}(a) and Prof. P. Vukusic (University of Exeter) for 
providing the picture of Papilio palinurus in our table of contents graphics entry.

\end{document}